\documentclass[useAMS,usenatbib]{mn2e}

\usepackage{graphicx}
\usepackage{rotating}

\title[Entropy profiles for a sample of X-ray galaxy groups and clusters]{A volume-limited sample of X-ray galaxy groups and clusters - I. Radial entropy and cooling time profiles}
\author[E. K. Panagoulia, A. C. Fabian \& J. S. Sanders]{E. K. Panagoulia$^{1}$\thanks{E-mail:
caillean@ast.cam.ac.uk}, A. C. Fabian$^{1}$ and J. S. Sanders$^{2}$\\
$^{1}$Institute of Astronomy, Madingley Road, Cambridge CB3 0HA\\
$^{2}$Max-Planck-Institute f\"{u}r extraterrestrische Physik, 85748, Garching, Germany.}
\begin{document}

\date{Accepted . Received ; in original form }
\pagerange{\pageref{firstpage}--\pageref{lastpage}} \pubyear{2013}

\maketitle

\label{firstpage}

\begin{abstract}
We present the first results of our study of a sample of 101 X-ray galaxy groups and clusters, which is volume-limited in each of three X-ray luminosity bins. The aim of this work is to study the properties of the innermost ICM in the cores of our groups and clusters, and to determine the effect of non-gravitational processes, such as active galactic nucleus (AGN) feedback, on the ICM. The entropy of the ICM is of special interest, as it bears the imprint of the thermal history of a cluster, and it also determines a cluster's global properties. Entropy profiles can therefore be used to examine any deviations from cluster self-similarity, as well as the effects of feedback on the ICM. We find that the entropy profiles are well-fitted by a simple powerlaw model, of the form $K(r) = \alpha\times(r/100 \rm{kpc})^{\beta}$, where $\alpha$ and $\beta$ are constants. We do not find evidence for the existence of an ``entropy floor'', i.e. our entropy profiles do not flatten out at small radii, as suggested by some previous studies. 
\end{abstract}


\section{Introduction}
While the general process of forming galaxy groups and clusters through hierarchical merging is now well-understood, details, such as the effect of feedback and radiative cooling on the cluster, are not. Early models of hierarchical structure formation included only the effects of gravitation, i.e. they were self-similar. The relations between observable cluster properties, such as X-ray luminosity and temperature, and how they scale with redshift and cluster mass, were quite specific \citep{Kaiser86, Kaiser91, Navarro95, Bryan98}. However, it soon became evident, from various observational studies, that clusters do not follow these predicted self-similar relations \citep[e.g.][]{Edge91, Allen98, Markevitch98}. Therefore, for theoretical predictions to match the observed scaling relations, non-gravitational effects, such as feedback and cooling in cluster cores, needed to be included in models \citep{Kaiser91, Voit02}.

The deviation of the observed scaling relations for clusters from self-similarity, implies the existence of processes which act to heat and cool the ICM. Cooling flows are one of these processes. For about 50\% of clusters, the core cooling time is shorter than the age of the cluster itself \citep[e.g.][]{Edge92, White97, Peres98}. Assuming that there are no heating processes at play in the core of these clusters, we would expect radiative cooling to result in a cooling flow, with a deposition rate of a few 100 M$\odot$ yr$^{-1}$ \citep[for a review, see][]{Fabian94}. However, this simple cooling flow model has trouble explaining the observed properties of groups and clusters. First of all, observations of the Fe L-line complex, which is a precise tool for measuring the amount of gas cooling at low temperatures, have revealed a much smaller amount of cooling gas with a temperature 2--3 times smaller than the virial temperature, than is predicted by the cooling flow model \citep{McNamara07, McNamara12, Peterson06}. In addition, cooling flows are expected to deposit cool gas in the centres of groups and clusters, so we would expect to see large amounts of molecular gas and high star formation rates towards cluster centres. Observations in the UV and optical indicate that star formation rates in clusters are only a small fraction of those predicted by the cooling flow model \citep[see e.g.][]{Nulsen87}, as is the amount of cold molecular gas observed in clusters. 

Although observed cluster properties are not explained solely by a simple cooling flow model, recent observations of brightest cluster galaxies (BCGs) indicate that the ICM does cool, albeit at a reduced rate. Some BCGs appear to harbour a large amount of cold gas, which is in the process of star formation \citep{Johnstone87, Allen95, Odea08, Odea10}. Additionally, many cool core (CC) clusters exhibit optical filaments, such as Abell~1795 \citep{Crawford05}. It is therefore clear that the ICM gas exists in several different phases, and across a very broad range of temperatures. As a result, a model more complex than the simple cooling flow model is needed to explain the full range of observed properties of galaxy groups and clusters. 

Both the breakdown of self-similarity and the cooling flow problem make it clear that one or more mechanisms are responsible for heating up the ICM in cluster cores, so preventing it from cooling and collapsing to form stars. Current models of the feedback process in galaxy clusters propose that the active galactic nucleus (AGN), that resides in the central galaxy of the cluster, is the main heating mechanism in cluster cores. Bursts of AGN activity deposit the necessary amount of energy into the surrounding ICM to reduce, and maybe even quench, radiative cooling \citep[for reviews, see][]{McNamara07, McNamara12, Fabian12}. Although the details of the AGN feedback loop are still poorly understood, there is no doubting the fact that AGN interact with the surrounding ICM, as reported in a number of studies \citep[see e.g.][]{Birzan04, Dunn08}. 

One of the most useful observables, that can be used to determine the effect of non-gravitational processes in clusters, is the entropy of the ICM. In studies of galaxy groups and clusters, the most commonly used definition for entropy is that of the specific entropy $ K = T\times n_{\rm e}^{-2/3}$, where $T$ is the temperature of the ICM gas, and $n_{\rm e}$ is the electron number density. By themselves, ICM temperature and electron number density cannot fully capture the thermal history of a cluster. The ICM temperature is a measure of the depth of the potential well of a cluster, while the ICM density reflects the potential well's ability to compress the ICM gas. However, $K$ can be used to study the thermal history of a cluster, as it is only susceptible to gains and losses of heat energy. This is because, if only gravitational effects determined cluster evolution, low-entropy gas would sink towards the centre of the cluster, while high-entropy gas would rise buoyantly towards larger radii. Hence, the radial entropy profiles of clusters would have the shape of a power law at larger radii (r $>$ 0.1r$_{200}$), with a constant, low-entropy core \citep{Voit05}. Large-scale deviations from this radial entropy distribution can then be used to examine how non-gravitational processes, such as AGN heating and radiative cooling, affect the ICM. Previous studies have found that the radial entropy profile in clusters flattens out at radii $<$ 0.1r$_{\rm virial}$, and that the dispersion in the core entropy is greater than that at large radii \citep[e.g.][]{David96, Ponman03, Pratt06, Cavagnolo09, McDonald13}. However, most of these studies were based on smaller and relatively specific samples. 

In this paper, we present the first results from our analysis of a volume- and X-ray luminosity-limited initial sample of 100 galaxy groups and clusters. The primary objective of this paper is to study the effect of AGN feedback and other heating processes at low redshifts, and determine the importance of each. We further seek to check whether this previously observed entropy floor exists at lower redshifts, and link this to what has been observed with samples of higher redshift clusters.

The paper is structured as follows: in Section 2, we discuss the selection criteria for our initial sample, and provide basic information on each of these objects. Data reduction is discussed in Section 3, while our deprojection spectral extraction methods are discussed in Section 3.1. 

In this paper, we adopt a flat ${\rm \Lambda}$CDM cosmology, with H$_{0}$ = 71 km s$^{-1}$ Mpc$^{-1}$, $\Omega_{\rm m}$ = 0.27 and $\Omega_{\Lambda}$ = 0.73. All abundances in this paper are relative to solar, as defined in \cite{Anders89}. In all the images shown in this paper, north is to the top and east is to the left. 

\section{A volume-limited sample of groups and clusters of galaxies}
The aim of this work is the study of the thermal properties of the ICM in groups and clusters of galaxies, and the determination of the importance and impact of non-gravitational processes, such as AGN feedback and conduction. We are particularly interested in the cores of clusters (i.e. the central $\sim$10 kpc), for the study of which high spatial and spectral resolution data are needed. To this end, we choose to look at a sample of nearby groups and clusters of galaxies, for which {\it Chandra} and/or {\it XMM-Newton} data are available. For this reason, a volume-limited sample of groups and clusters was constructed using the Northern {\it ROSAT} All Sky catalogue \citep[NORAS;][]{Bohringer00}, and the {\it ROSAT-ESO} Flux Limited X-ray galaxy cluster survey \citep[REFLEX;][]{Bohringer04}. The sources were selected to lie at a maximum distance of 300 Mpc from us, as listed in the NORAS and REFLEX catalogues, which corresponds to a redshift of 0.071 in our cosmology. The number of sources meeting this criterion was 105 from the NORAS catalogue, and 167 sources from the REFLEX catalogue. An additional 17 sources from the NORAS catalogue, that were listed as promising candidates and were later identified as previously listed galaxy clusters by \cite{Bohringer00} after a re-examination of the {\it ROSAT} All-Sky Survey II (RASS II) catalogue, were also included in our initial sample. Sources that were listed as having a redshift and/ or X-ray luminosity of zero in the NORAS or REFLEX catalogues were not considered during the initial sample selection. We then checked the {\it Chandra} archive for data on each of the sources. If {\it Chandra} data were not available, {\it XMM-Newton} data were used, where available. {\it Chandra} data were preferred in order to use the greater spatial resolution and smaller point-spread function (PSF) of the ACIS detectors, in order to facilitate the study of the inner boundaries of the ICM and any extended structure in more detail. The ACIS instruments also have much lower background levels compared to the EPIC instruments on board {\it XMM-Newton}, which is important when studying diffuse sources, such as galaxy groups and clusters.

\begin{figure*}
  \includegraphics[trim = 2.0cm 3.0cm 1.0cm 0.0cm, clip, height=7.5in, width=5in, angle=270]{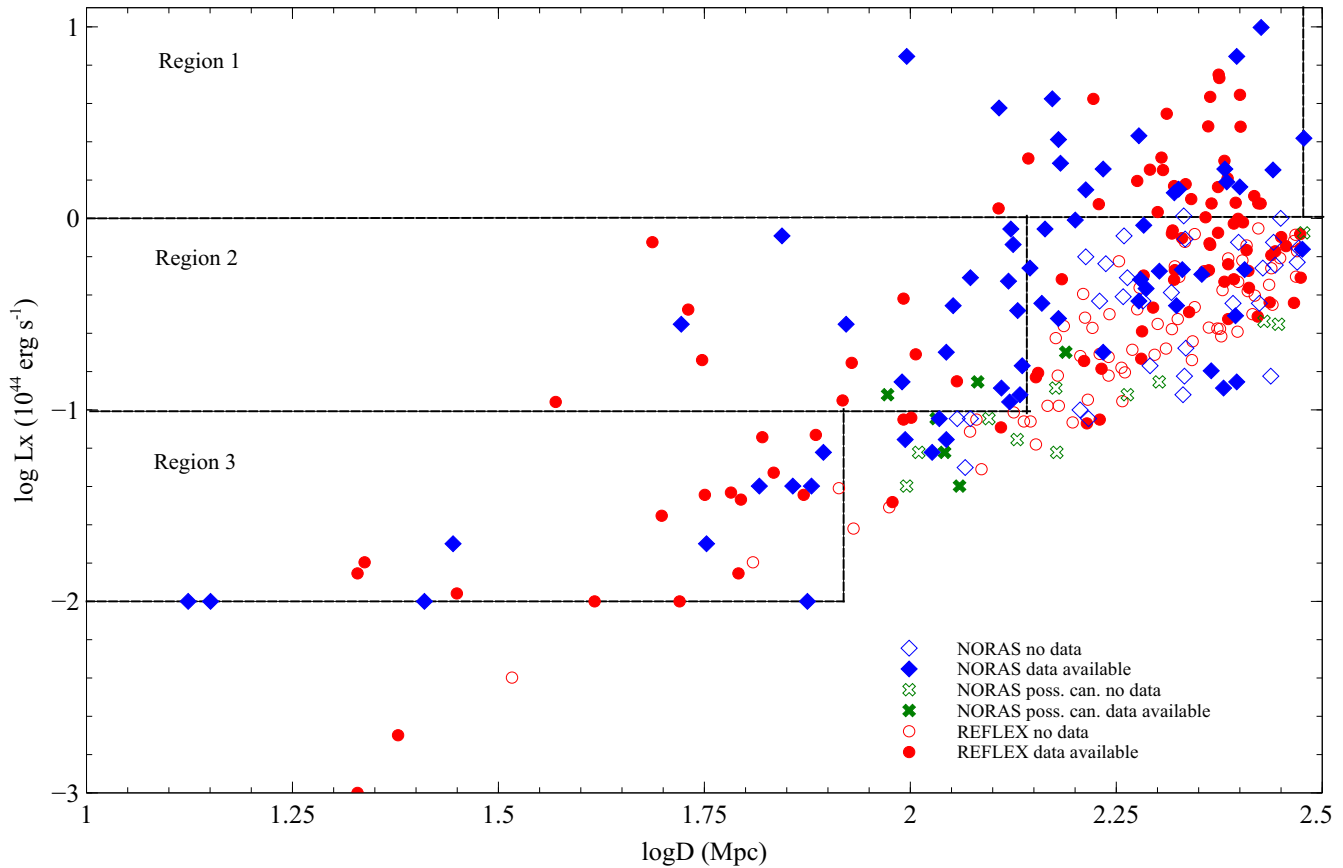}
  \caption[]{The initial volume-limited sample of 289 galaxy groups and clusters. The regions corresponding to different cuts in L$_{\rm X}$ and distance are also shown (see Section 2).}
  \label{fig:sample}
  \end{figure*}

Figure \ref{fig:sample} shows the total 289 sources from our initial volume-limited sample. The red circles correspond to REFLEX sources, the blue diamonds to NORAS sources, and the green crosses to the 17 NORAS sources that were identified as galaxy clusters after the RASS II galaxy catalogue was re-examined. The filled-in symbols correspond to sources for which {\it Chandra} and/ or {\it XMM-Newton} data are available, while the empty symbols correspond to sources for which neither {\it XMM-Newton} or {\it Chandra} are available. This figure clearly demonstrates that there is a very noticeable lack of data for clusters with $L_{\rm X}$ = 10$^{43}$ --10$^{44}$ erg s$^{-1}$, which are at a distance $>$140 Mpc. Taking this fact into consideration, in order to produce a sample that is as statistically complete as possible, cuts in $L_{\rm X}$ and distance were made, so as to reduce the number of clusters in our sample with no data. Our final sample consists of the following ``regions'' of Figure \ref{fig:sample}: 
\begin{itemize}
  \item Region 3 corresponds to clusters with $L_{\rm X}$ = 10$^{42}$--10$^{43}$ erg s$^{-1}$ at a distance $\leq$83 Mpc. There are 16 sources within these limits from the REFLEX catalogue (two of which have no data), and 10 sources from the NORAS catalogue (all of which have data). 
    \item Region 2 encompasses all clusters with $L_{\rm X}$ = 10$^{43}$--10$^{44}$ erg s$^{-1}$, which lie at a distance $\leq$140 Mpc. Region 2 consists of 10 REFLEX sources (all of which have data), and 18 NORAS sources, including 2 sources that were confirmed as clusters by \cite{Bohringer00} (all sources have data). 
      \item Region 1 comprises all sources with $L_{\rm X}$$ > $ 10$^{44}$ erg s$^{-1}$, which are at a distance $\leq$300 Mpc. There are 28 REFLEX clusters (all of which have data) and 19 NORAS clusters (two of which have no data) that meet this criterion. 
\end{itemize}

In total, our volume- and $L_{\rm X}$-limited sample consists of 101 galaxy groups and clusters, for only 4 of which there are no {\it Chandra} or {\it XMM-Newton} data. We note that, for some of the sources for which data are available, the observation exposures were too short, resulting in data of poor quality that could not be used in spectral analysis. Tables \ref{tab:reflexsources} and \ref{tab:norassources} show the list of REFLEX and NORAS sources, respectively, that were included in our final sample. The sources in bold are the sources for which no {\it Chandra} or {\it XMM-Newton} observations were available. The two sources in italics in Table \ref{tab:norassources} are the two NORAS sources that were confirmed as galaxy clusters by \cite{Bohringer00}. The sources whose names are underlined in the tables were the ones for which we used {\it XMM-Newton} data in our analysis. The starred sources are the ones for which we were able to obtain radial deprojected electron number density and temperature profiles, and hence were used in the creation of the overall profiles described further on.

\begin{table*}
%\caption{List of REFLEX sources in our catalogue.}
\begin{center}
\footnotesize{ 
\begin{tabular}{cccccccccc} 
    \multicolumn{1}{c}{Source name}&\multicolumn{1}{c}{Alt. source name}&\multicolumn{3}{c}{RA (2000)}&\multicolumn{3}{c}{DEC (2000)}&\multicolumn{1}{c}{Redshift}&\multicolumn{1}{c}{$L_{\rm {X}}$ ($\times$ 10$^{44}$ erg s$^{-1}$)} \\ 
    \multicolumn{1}{c}{(1)}&\multicolumn{1}{c}{(2)}&\multicolumn{3}{c}{(3)}&\multicolumn{3}{c}{(4)}&\multicolumn{1}{c}{(5)}&\multicolumn{1}{c}{(6)} \\
    \hline
\underline{RXCJ1336.6-3357}* & A3565 & 13 & 36 & 38.8 & -33 & 57 & 30 & 0.0123 & 0.008 \\
RXCJ2009.9-4823* & S0851 & 20 & 09 & 54.1 & -48 & 23 & 35 & 0.0097 & 0.007 \\
RXCJ1506.4+0136* & NGC5846 & 15 & 06 & 29.7 & +01 & 36 & 08 & 0.0066 & 0.008 \\
\underline{RXCJ1347.2-3025} & A3574W & 13 & 47 & 12.3 & -30 & 25 & 10 & 0.0145 & 0.012 \\
RXCJ1501.1+0141* & NGC5813 & 15 & 01 & 11.9 & +01 & 41 & 53 & 0.0050 & 0.007 \\
RXCJ0338.4-3526* & FORNAX & 03 & 38 & 27.9 & -35 & 26 & 54 & 0.0051 & 0.012 \\
{\bf RXCJ1257.1-1339} & - & 12 & 57 & 10.1 & -13 & 39 & 20 & 0.0151 & 0.012 \\
{\bf RXCJ2018.4-4102} & IC4992 & 20 & 18 & 25.6 & -41 & 02 & 48 & 0.0192 & 0.032 \\
RXCJ1304.2-3030 & - & 13 & 04 & 16.7 & -30 & 30 & 55 & 0.0117 & 0.025 \\
RXCJ1253.0-0912* & HCG62 & 12 & 53 & 05.5 & -09 & 12 & 01 & 0.0146 & 0.037 \\
\underline{RXCJ1847.3-6320} & S0805 & 18 & 47 & 20.0 & -63 & 20 & 13 & 0.0146 & 0.029 \\
RXCJ0125.5+0145* & NGC533 & 01 & 25 & 30.2 & +01 & 45 & 44 & 0.0174 & 0.032 \\
RXCJ1403.5-3359 & S0753 & 14 & 03 & 35.9 & -33 & 59 & 16 & 0.0132 & 0.032 \\
\underline{RXCJ1349.3-3018} & A3574E & 13 & 49 & 19.3 & -30 & 18 & 34 & 0.0160 & 0.043 \\
RXCJ1050.4-1250* & USGCS152 & 10 & 50 & 25.5 & -12 & 50 & 47 & 0.0155 & 0.059 \\
RXCJ0125.6-0124 & A0194 & 01 & 25 & 40.8 & -01 & 24 & 26 & 0.0180 & 0.070 \\
RXCJ1315.3-1623* & NGC5044 & 13 & 15 & 24.0 & -16 & 23 & 23 & 0.0087 & 0.097 \\
\underline{RXCJ1840.6-7709}* & - & 18 & 40 & 37.2 & -77 & 09 & 20 & 0.0194 & 0.087 \\
\underline{RXCJ2315.7-0222}* & - & 23 & 15 & 45.2 & -02 & 22 & 37 & 0.0267 & 0.134 \\
\underline{RXCJ0624.6-3720} & A3390 & 06 & 24 & 36.7 & -37 & 20 & 09 & 0.0333 & 0.142 \\
RXCJ1204.4+0154 & MKW4 & 12 & 04 & 25.2 & +01 & 54 & 02 & 0.0199 & 0.153 \\
RXCJ0419.6+0224* & NGC1550 & 04 & 19 & 37.8 & +02 & 24 & 50 & 0.0131 & 0.153 \\
RXCJ0110.0-4555 & A2877 & 01 & 10 & 00.4 & -45 & 55 & 22 & 0.0238 & 0.179 \\
RXCJ1036.6-2731* & A1060 & 10 & 36 & 41.8 & -27 & 31 & 28 & 0.0126 & 0.297 \\
RXCJ1407.4-2700* & A3581 & 14 & 07 & 28.1 & -27 & 00 & 55 & 0.0230 & 0.316 \\
RXCJ1248.7-4118* & A3526 & 12 & 48 & 47.9 & -41 & 18 & 28 & 0.0114 & 0.721 \\
RXCJ1252.5-3116* & - & 12 & 52 & 34.1 & -31 & 16 & 04 & 0.0535 & 0.861 \\
\underline{RXCJ0601.7-3959}* & A3376 & 06 & 01 & 45.7 & -39 & 59 & 34 & 0.0468 & 1.036 \\
RXCJ2347.7-2808* & A4038 & 23 & 47 & 43.2 & -28 & 08 & 29 & 0.0300 & 1.014 \\
RXCJ0425.8-0833* & RBS0540 & 04 & 25 & 51.4 & -08 & 33 & 33 & 0.0397 & 1.008 \\
RXCJ0330.0-5235 & A3128 & 03 & 30 & 00.7 & -52 & 35 & 46 & 0.0624 & 1.171 \\
RXCJ1254.6-2913* & A3528S & 12 & 54 & 41.4 & -29 & 13 & 24 & 0.0544 & 1.064 \\
RXCJ0011.3-2851* & A2734 & 00 & 11 & 20.7 & -28 & 51 & 18 & 0.0620 & 1.089 \\
RXCJ2205.6-0535 & A2415 & 22 & 05 & 40.5 & -05 & 35 & 36 & 0.0582 & 1.135 \\
\underline{RXCJ0626.3-5341}* & A3391 & 06 & 26 & 22.8 & -53 & 41 & 44 & 0.0514 & 1.198 \\
\underline{RXCJ0351.1-8212} & S0405 & 03 & 51 & 08.9 & -82 & 12 & 60 & 0.0613 & 1.243 \\
RXCJ1257.2-3022 & A3532 & 12 & 57 & 16.9 & -30 & 22 & 37 & 0.0554 & 1.340 \\
RXCJ1333.6-3139* & A3562 & 13 & 33 & 36.3 & -31 & 39 & 40 & 0.0490 & 1.357 \\
\underline{RXCJ0627.2-5428}* & A3395 & 06 & 27 & 14.4 & -54 & 28 & 12 & 0.0506 & 1.403 \\
RXCJ0056.3-0112 & A0119 & 00 & 56 & 18.3 & -01 & 12 & 60 & 0.0442 & 1.505 \\
RXCJ0102.7-2152* & A0133 & 01 & 02 & 42.1 & -21 & 52 & 25 & 0.0569 & 1.439 \\
RXCJ2357.0-3445* & A4059 & 23 & 57 & 02.3 & -34 & 45 & 38 & 0.0475 & 1.698 \\
\underline{RXCJ1326.9-2710}* & A1736 & 13 & 26 & 54.0 & -27 & 10 & 60 & 0.0458 & 1.778 \\
RXCJ2313.9-4244* & S1101 & 23 & 13 & 58.6 & -42 & 44 & 02 & 0.0564 & 1.738 \\
RXCJ0433.6-1315* & A0496 & 04 & 33 & 38.4 & -13 & 15 & 33 & 0.0326 & 1.746 \\
RXCJ1257.1-1724* & A1644 & 12 & 57 & 09.8 & -17 & 24 & 01 & 0.0473 & 1.952 \\
RXCJ0342.8-5338* & A3158 & 03 & 42 & 53.9 & -53 & 38 & 07 & 0.0590 & 2.951 \\
RXCJ0918.1-1205* & A0780 & 09 & 18 & 06.5 & -12 & 05 & 36 & 0.0539 & 2.659 \\
RXCJ1327.9-3130 & A3558 & 13 & 27 & 57.5 & -31 & 30 & 09 & 0.0480 & 3.096 \\
RXCJ1347.4-3250* & A3571 & 13 & 47 & 28.4 & -32 & 50 & 59 & 0.0391 & 3.996 \\
RXCJ0909.1-0939* & A0754 & 09 & 09 & 08.4 & -09 & 39 & 58 & 0.0542 & 3.879 \\
RXCJ0431.4-6126 & A3266 & 04 & 31 & 24.1 & -61 & 26 & 38 & 0.0589 & 4.019 \\
RXCJ2012.5-5649* & A3667 & 20 & 12 & 30.5 & -56 & 49 & 55 & 0.0556 & 5.081 \\
RXCJ0041.8-0918* & A0085 & 00 & 41 & 50.1 & -09 & 18 & 07 & 0.0555 & 5.293 \\
    \hline
\end{tabular}
}
\end{center}
\caption{List of REFLEX sources in our catalogue. (1) source name as in the REFLEX catalogue, (2) alternative source name (sometimes refering to the central dominant galaxy), (3) and (4) source right ascension and declination in epoch 2000 coordinates, (5) source redshift, (6) source X-ray luminosity in the rest frame 0.1--2.4 keV band.}
 \label{tab:reflexsources}
%\end{center}
\end{table*}

\begin{table*}
%\caption{List of NORAS sources in our catalogue.}
\label{tab:reflexobs}
\centering
\footnotesize{ 
\begin{tabular}{ccc}
\multicolumn{1}{c}{Source name}&\multicolumn{1}{c}{Obs IDs}&\multicolumn{1}{c}{Clean exposure time} \\ 
\multicolumn{1}{c}{(1)}&\multicolumn{1}{c}{(2)}&\multicolumn{1}{c}{(3)} \\
    \hline
\underline{RXCJ1336.6-3357}* & 0672870101 & 41.2 ks, 40.2 ks, 34.5 ks\\ 
RXCJ2009.9-4823* &  3191, 11753 & 95.5 ks \\
RXCJ1506.4+0136* & 788, 7923 & 112.3 ks  \\
\underline{RXCJ1347.2-3025} & 0101040401 & - - -\\
RXCJ1501.1+0141* & 12951, 12952, 13246, 13247, 13253, 13255, 5907, 9517 & 604.2 ks \\
RXCJ0338.4-3526* & 319, 4172, 4174, 9530, 9798 & 218.6 ks \\
{\bf RXCJ1257.1-1339} & - & - \\
{\bf RXCJ2018.4-4102} & - & - \\
RXCJ1304.2-3030 & 4997, 4998 & 28.9 ks \\  
RXCJ1253.0-0912* & 10462, 10874, 921 & 166.2 ks \\
\underline{RXCJ1847.3-6320} & 0405550401 & 26.5 ks \\ 
RXCJ0125.5+0145* & 2880 & 33.9 ks\\
RXCJ1403.5-3359 & 4999, 5000 & 29.0 ks \\
\underline{RXCJ1349.3-3018} & 0101040401 & - - 9.2 ks \\
RXCJ1050.4-1250* & 3243 & 26.0 ks \\
RXCJ0125.6-0124 & 3c40 & \\  
RXCJ1315.3-1623* & 3225, 3664, 9399 & 222.0 ks \\
\underline{RXCJ1840.6-7709}* & 0405550301 & 15.0 ks \\
\underline{RXCJ2315.7-0222}* & 0501110101 & 33.4 ks \\
\underline{RXCJ0624.6-3720} & 0151270401 & 5.8 ks \\
RXCJ1204.4+0154 & 3234 & 27.4 ks \\
RXCJ0419.6+0224* & 5800, 5801 & 88.5 ks \\
RXCJ0110.0-4555 & 4971 & 23.4 ks \\
RXCJ1036.6-2731* & 2220 & 30.6 ks \\
RXCJ1407.4-2700* & 12884 & 84.5 ks \\
RXCJ1248.7-4118* & 4954, 4955, 504, 5310 & 198.0 ks \\
RXCJ1252.5-3116* & 12275 & 9.9 ks \\  
\underline{RXCJ0601.7-3959}* & 0504140101 & 42.7 ks \\
RXCJ2347.7-2808* & 4992 & 33.1 ks \\
RXCJ0425.8-0833* & 4183 & 10.0 ks \\
RXCJ0330.0-5235 & 893 & 19.6 ks \\
RXCJ1254.6-2913* & 10746 & 10.0 ks \\
RXCJ0011.3-2851* & 5797 & 19.9 ks \\
RXCJ2205.6-0535 & 12272 & 9.9 ks \\
\underline{RXCJ0626.3-5341}* & 0505210401 & 24.9 ks\\
\underline{RXCJ0351.1-8212} & 0675471101 & 7.2 ks \\
RXCJ1257.2-3022 & 10745 & 9.6 ks \\
RXCJ1333.6-3139*  & 4167 & 19.1 ks \\
\underline{RXCJ0627.2-5428}* & 0400010301 & 29.1 ks \\
RXCJ0056.3-0112 & 4180, 7918 & 56.4 ks \\
RXCJ0102.7-2152* & 2203, 9897 & 102.9 ks \\
RXCJ2357.0-3445* & 897, 5785 & 107.1 ks \\
\underline{RXCJ1326.9-2710}* & 0300210301, 0402190101, 0505210201 & 7.7 ks 8.1 ks - / 8.6 ks 7.9 ks - / 11.2 ks 12.2 ks - \\
RXCJ2313.9-4244* & 11758 & 96.5 ks \\ 
RXCJ0433.6-1315* & 4976 & 53.2 ks \\
RXCJ1257.1-1724* & 2206, 7922 & 69.9 ks \\
RXCJ0342.8-5338* & 3201, 3712 & 55.7 ks \\
RXCJ0918.1-1205* & 4969, 4970 & 165.9 ks \\
RXCJ1327.9-3130 & 1646 & 8.3 ks \\
RXCJ1347.4-3250* & 4203 & 22.3 ks \\
RXCJ0909.1-0939* & 577 & 43.6 ks \\
RXCJ0431.4-6126 & 899 & 29.4 ks \\
RXCJ2012.5-5649* & 5751, 5752, 5753, 6292, 6295, 6296, 889 & 433.0 ks \\
RXCJ0041.8-0918* & 904 & 38.4 ks \\
    \hline
\end{tabular}
}
\caption{List of Obs IDs and clean exposure times for the REFLEX sources in our catalogue. (1) source name as in the REFLEX catalogue, (2) {\it Chandra} or {\it XMM-Newton} Obs IDs used in the analysis, (3) total clean exposure time. In the case of {\it XMM-Newton} data, we show the MOS1, MOS2 and pn clean exposure times separately.} 
%\end{center}
\end{table*}

\begin{table*}
%\caption{List of NORAS sources in our catalogue.}
\begin{center}
\footnotesize{ 
\begin{tabular}{ccccccc} 
    \multicolumn{2}{c}{Source name}&\multicolumn{1}{c}{Alt. source name}&\multicolumn{1}{c}{RA (2000)}&\multicolumn{1}{c}{DEC (2000)}&\multicolumn{1}{c}{Redshift}&\multicolumn{1}{c}{$L_{\rm {X}}$ ($\times$ 10$^{44}$ erg s$^{-1}$)} \\ 
    \multicolumn{2}{c}{(1)}&\multicolumn{1}{c}{(2)}&\multicolumn{1}{c}{(3)}&\multicolumn{1}{c}{(4)}&\multicolumn{1}{c}{(5)}&\multicolumn{1}{c}{(6)} \\
    \hline
RXC & J1242.8+0241* & NGC4636 & 190.7063 & 2.6856 & 0.0031 & 0.01 \\
RXC & J1229.7+0759* & M49-Virgo & 187.4403 & 7.9870 & 0.0033 & 0.01 \\
RXC & J1506.4+0136* & NGC5846 & 226.6241 & 1.6049 & 0.0060 & 0.01 \\
RXC & J1501.2+0141* & NGC5813 & 225.3016 & 1.6960 & 0.0065 & 0.02 \\
RXC & J1751.7+2304* & NGC6482 & 267.9480 & 23.0705 & 0.0132 & 0.02 \\
RXC & J0123.1+3327* & - & 20.7970 & 33.4620 & 0.0153 & 0.04 \\
RXC & J0200.2+3126* & NGC0777 & 30.0687 & 31.4365 & 0.0168 & 0.04 \\
\underline{RXC} & \underline{J0920.0+0102} & MKW1s & 140.0020 & 1.0401 & 0.0175 & 0.01 \\
\underline{RXC} & \underline{J0110.9+3308} & NGC0410 & 17.7419 & 33.1494 & 0.0177 & 0.04 \\
RXC & J0125.5+0145* & N0533 & 21.3744 & 1.7629 & 0.0183 & 0.06 \\
RXC & J0419.6+0224* & N1550 & 64.9067 & 2.4144 & 0.0123 & 0.28 \\
RXC & J0152.7+3609* & A0262 & 28.1948 & 36.1513 & 0.0163 & 0.81 \\
RXC & J1204.4+0154* & MKW4 & 181.1065 & 1.9010 & 0.0195 & 0.28 \\
\underline{RXC} & \underline{J0751.3+5012} & UGC04052 & 117.8437 & 50.2125 & 0.0228 & 0.14 \\
RXC & J1223.1+1037* & NGC4325 & 185.7772 & 10.6240 & 0.0258 & 0.20 \\
RXC & J1440.6+0328* & MKW8 & 220.1592 & 3.4765 & 0.0263 & 0.35 \\
RXC & J1715.3+5724* & NGC6338 & 258.8414 & 57.4074 & 0.0276 & 0.49 \\
RXC & J1628.6+3932* & A2199 & 247.1582 & 39.5487 & 0.0299 & 3.77 \\
\underline{RXC} & \underline{J1627.6+4055} & A2197 & 246.9173 & 40.9197 & 0.0301 & 0.13 \\
RXC & J1733.0+4345* & IC1262 & 263.2607 & 43.7629 & 0.0307 & 0.47 \\
RXC & J1615.5+1927* & NGC6098 & 243.8947 & 19.4600 & 0.0308 & 0.11 \\
RXC & J2338.4+2659 & A2634 & 354.6071 & 27.0126 & 0.0309 & 0.88 \\
RXC & J0341.2+1524* & - & 55.3206 & 15.4074 & 0.0311 & 0.73 \\
RXC & J1617.5+3458 & NGC6107 & 244.3635 & 34.9367 & 0.0315 & 0.33 \\
\underline{RXC} & \underline{J1627.3+4240}* & A2192 & 246.8482 & 42.6784 & 0.0317 & 0.12 \\
RXC & J1109.7+2145 & A1177 & 167.4294 & 21.7620 & 0.0319 & 0.17 \\
RXC & J1604.9+2355* & AWM4 & 241.2377 & 23.9206 & 0.0326 & 0.55 \\
RXC & J1259.7+2756* & Coma & 194.9294 & 27.9386 & 0.0231 & 7.01 \\
RXC & J1628.6+3932* & A2199 & 247.1582 & 39.5487 & 0.0299 & 3.77 \\
RXC & J0338.6+0958* & 2A0335 & 54.6699 & 9.9745 & 0.0347 & 4.21 \\
RXC & J1516.7+0701* & A2052 & 229.1834 & 7.0185 & 0.0353 & 2.58 \\
RXC & J1523.0+0836* & A2063 & 230.7724 & 8.6025 & 0.0355 & 1.94 \\
RXC & J0721.3+5547* & A0576 & 110.3426 & 55.7864 & 0.0381 & 1.41 \\
RXC & J2344.9+0911 & A2657 & 356.2376 & 9.1980 & 0.0400 & 1.81 \\
RXC & J1521.8+0742* & MKW3s & 230.4583 & 7.7088 & 0.0442 & 2.70 \\
RXC & J0246.0+3653* & A0376 & 41.5108 & 36.8879 & 0.0488 & 1.36 \\
RXC & J2113.8+0233 & IC1365 & 318.4745 & 2.5555 & 0.0494 & 1.42 \\
{\bf RXC} & {\bf J1811.0+4954} & - & 272.7503 & 49.9110 & 0.0501 & 1.03 \\
RXC & J2350.8+0609* & A2665 & 357.7109 & 6.1611 & 0.0562 & 1.81 \\
RXC & J2336.5+2108* & A2626 & 354.1261 & 21.1424 & 0.0565 & 1.55 \\
RXC & J1703.8+7838 & A2256 & 255.9532 & 78.6443 & 0.0581 & 7.01 \\
RXC & J1454.5+1838* & A1991 & 223.6309 & 18.6420 & 0.0586 & 1.46 \\
RXC & J1348.8+2635* & A1795 & 207.2207 & 26.5956 & 0.0622 & 9.93 \\
RXC & J1303.7+1916* & A1668 & 195.9398 & 19.2715 & 0.0643 & 1.79 \\
{\bf RXC} & {\bf J0209.5+1946} & A0311 & 32.3796 & 19.7695 & 0.0657 & 1.00\\
RXC & J1336.1+5912 & A1767 & 204.0255 & 59.2079 & 0.0701 & 2.62 \\
{\it RXC} & {\it J1329.5+1147} & MKW11 & 202.3836 & 11.7892 & 0.0220 & 0.12\\
{\it RXC} & {\it J1206.6+2811} & NGC4104 & 181.6560 & 28.1835 & 0.0283 & 0.14\\
    \hline
\end{tabular}
}
\end{center}
\caption{List of NORAS sources in our catalogue. (1) source name as in the NORAS catalogue, (2) alternative source name (sometimes refering to the central dominant galaxy), (3) and (4) source right ascension and declination in epoch 2000 coordinates, (5) source redshift, (6) source X-ray luminosity in the rest frame 0.1--2.4 keV band.}
 \label{tab:norassources}
%\end{center}
\end{table*}

\begin{table*}
%\caption{List of NORAS sources in our catalogue.}
%\label{tab:norasobs}
\centering
\footnotesize{ 
\begin{tabular}{cccc}
\multicolumn{2}{c}{Source name}&\multicolumn{1}{c}{Obs IDs}&\multicolumn{1}{c}{Clean exposure time} \\ 
\multicolumn{2}{c}{(1)}&\multicolumn{1}{c}{(2)}&\multicolumn{1}{c}{(3)} \\
    \hline
RXC & J1242.8+0241*	      &        323, 3925, 4415 & 188.9 ks \\
RXC & J1229.7+0759*	      &        12888, 12889 & 293.1 ks \\
RXC & J1506.4+0136*	      &        788, 7923 & 112.3 ks \\
RXC & J1501.2+0141*	      &        12951, 12952, 13246, 13247, 13253, 13255, 5907, 9517 & 604.2 ks \\
RXC & J1751.7+2304*	      &        3218 & 18.4 ks \\
RXC & J0123.1+3327*	      &        10536 & 18.4 ks \\
RXC & J0200.2+3126*	      &        5001 & 10.0 ks \\
\underline{RXC} & \underline{J0920.0+0102}	      &     0673180201    & 19.5 ks 19.7 ks 11.2 ks \\
\underline{RXC} & \underline{J0110.9+3308}	      &        0203610201 & 16.9 ks 16.8 ks 13.4 ks \\
RXC & J0125.5+0145*	      &        2880 & 33.9 ks \\
RXC & J0419.6+0224*	      &        5800, 5801 & 88.5 ks \\
RXC & J0152.7+3609*	      &        2215, 7921 & 138.6 ks \\
RXC & J1204.4+0154*	      &        3234 & 27.4 ks \\
\underline{RXC} & \underline{J0751.3+5012}	      &        0151270201 & 9.3 ks 9.4 ks - \\
RXC & J1223.1+1037*	      &        3232 & 26.2 ks \\
RXC & J1440.6+0328*	      &        4942 & 23.1 ks \\
RXC & J1715.3+5724*	      &        4194 & 45.9 ks \\
RXC & J1628.6+3932*	      &        10748, 10803, 10804, 10805 & 119.7 ks \\
\underline{RXC} & \underline{J1627.6+4055}	      &        0203710101 & - - - \\
RXC & J1733.0+4345*	      &        2018, 6949, 7321, 7322 & 137.8 ks \\
RXC & J1615.5+1927*	      &        10230 & 44.2 ks \\
RXC & J2338.4+2659	      &        4816 & 48.9 ks \\
RXC & J0341.2+1524*	      &        4182 & 23.5 ks \\
RXC & J1617.5+3458	      &        8180 & 19.3 ks \\
\underline{RXC} & \underline{J1627.3+4240}*	      &        0670350501 & 20.6 ks 20.6 ks 19.0 ks \\
RXC & J1109.7+2145	      &        6940 & 32.1 ks\\
RXC & J1604.9+2355*	      &        9423 & 74.3 ks\\
RXC & J1259.7+2756*	      &        13993, 13994, 13995, 13996, 14406, 14410, 14411, 14415, 9714 & 535.4 ks\\
RXC & J1628.6+3932*	      &        10748, 10803, 10804, 10805 & 119.7 ks\\
RXC & J0338.6+0958*	      &        919, 7939, 9792 & 102.0 ks\\
RXC & J1516.7+0701*	      &        5807, 10477, 10478, 10479, 10480, 10879, 10914, 10916, 10917 & 599.7 ks \\
RXC & J1523.0+0836*	      &        5795, 6263 & 26.1 ks \\
RXC & J0721.3+5547*	      &        3289 & 29.1 ks \\
RXC & J2344.9+0911	      &        4941 & 16.0 ks \\
RXC & J1521.8+0742*	      &        900 & 57.3 ks \\
RXC & J0246.0+3653*	      &        12277 & 10.4 ks \\
RXC & J2113.8+0233	      &        10747 & 10.0 ks \\
{\bf RXC} & {\bf J1811.0+4954}	&      - & - \\
RXC & J2350.8+0609*	      &        12280 & 9.9 ks \\
RXC & J2336.5+2108*	      &        3192 & 24.4 ks \\
RXC & J1703.8+7838	      &        1386, 2419 & 20.7 ks \\
RXC & J1454.5+1838*	      &        3193 & 37.3 ks \\
RXC & J1348.8+2635*	      &        493, 494 & 37.1 ks \\
RXC & J1303.7+1916*	      &        12877 & 9.0 ks \\
{\bf RXC} & {\bf J0209.5+1946} &	      - & - \\
RXC & J1336.1+5912	        &      12282 & 9.9 ks \\
{\it RXC} & {\it J1329.5+1147}	&       3216 & 33.5 ks \\
{\it RXC} & {\it J1206.6+2811}	&       6939 & 35.9 ks \\
    \hline
\end{tabular}
}
\caption{List of Obs IDs and clean exposure times for the NORAS sources in our catalogue. (1) source name as in the NORAS catalogue, (2) {\it Chandra} or {\it XMM-Newton} Obs IDs used in the analysis, (3) total clean exposure time. In the case of {\it XMM-Newton} data, we show the MOS1, MOS2 and pn clean exposure times separately.} 
\label{tab:norasobs}
%\end{center}
\end{table*}

\section{Observations and data preparation}
\subsection{{\it\textbf{Chandra}} data}
The observation IDs (Obs IDs) for each cluster, which were used in our analysis, are shown in Column 2 of Tables \ref{tab:reflexobs} and \ref{tab:norasobs}. New Level 2 event files were created from the Level 1 event files, using the {\sc ciao} {\sc acis\_reprocess\_events} pipeline, so that the most recent calibration files were applied (as of {\sc ciao} version 4.4). These include the appropriate charge transfer inefficiency (CTI) and gain correction files. Background lightcurves were created for each dataset, by selecting an appropriate chip that was used in the observation, in order to get rid of periods of X-ray background flaring. If the ACIS-S detector was used, the ACIS-S1 chip was chosen by default to provide a background lightcurve. If this chip was not active during the observation, the ACIS-S3 chip was used instead. In both these cases, the lightcurves were created in the 2.5--7.0 keV energy band. In all other cases, the chips ACIS-I0, ACIS-I1 and ACIS-I2 were used to obtain background lightcurves in the 0.5--12.0 keV energy band. These lightcurves were then visually examined for periods of background flaring, which were then excluded from any subsequent analysis. The total clean time available for each source, after the periods of X-ray background flaring had been excised, is shown in Column 3 of Tables \ref{tab:reflexobs} and \ref{tab:norasobs}. For the sources for which {\it XMM-Newton} data were used in the analysis, we give the MOS1, MOS2 and pn clean exposure times, on the left, central and right section of Column 3, respectively. If multiple datasets were available for one source, they were deprojected onto the same set of coordinates. 

Blank-sky observations were selected and adjusted to match the individual observations, and were used to provide background images and spectra. To account for the increase in background levels with time during the blank-sky observations, their exposure times were adjusted so that their count rate in the 9.0--12.0 keV energy band matched that of the corresponding cluster observation. The blank-sky datasets were then also reprojected onto the same set of coordinates as their corresponding source observations. If more than one dataset was available for each source, in order to create a total background image, the exposure time of each reprojected background image was weighed by the ratio of the corresponding cluster observation exposure time over the total cluster observation time. The individual weighted, reprojected background images were then added together to create a total background image. 

A 0.5--7.0 keV background-subtracted, exposure-corrected composite image of each cluster was created. The {\sc ciao} {\sc wavdetect} routine was used to identify point sources in each of these images. The point sources were then visually verified, and excluded from all subsequent spectral analysis. 
 
\subsection{{\it\textbf{XMM-Newton}} data}

Data were used from all three EPIC detectors on board {\it XMM-Newton}. The MOS and pn data were reprocessed using the {\sc emchain} and {\sc epchain} commands, respectively, to use the appropriate calibration files, as of {\sc sas} version 12.0.0. In order to filter out periods of background flaring, lightcurves were extracted for each of the three EPIC detectors. To this end, each event file was filtered using {\sc pattern==0} and {\sc \#xmmea\_em} or {\sc \#xmmea\_ep} for the MOS and pn detectors, respectively. Lightcurves were then extracted for each of the three detectors, in the $\geq$10 keV band for the MOS detectors, and in the 10--12 keV band for the pn detector, in order to screen out periods of X-ray background flaring. These lightcurves were binned using appropriate time bins, and visually inspected for periods of background flaring, which were excluded from any subsequent analysis. We give the MOS1, MOS2 and pn clean exposure times, on the left, central and right section of Column 3, respectively, in Tables \ref{tab:reflexobs} and \ref{tab:norasobs}. 0.5--7.0 keV exposure-corrected images were then created, by adding together the individual images from each of the detectors, and were visually examined for point sources. These point sources were then excised from any further spectral analysis. 

%A region free from group or cluster emission, and at a similar distance off-axis as the object of interest, was chosed to provide a local background during spectral analysis. 

\section{Spectral analysis and results}
\subsection{Spectral analysis}

\begin{figure*}
\includegraphics[trim=0cm 14cm 4.5cm 0cm, clip, height=3.4in, width=3.4in]{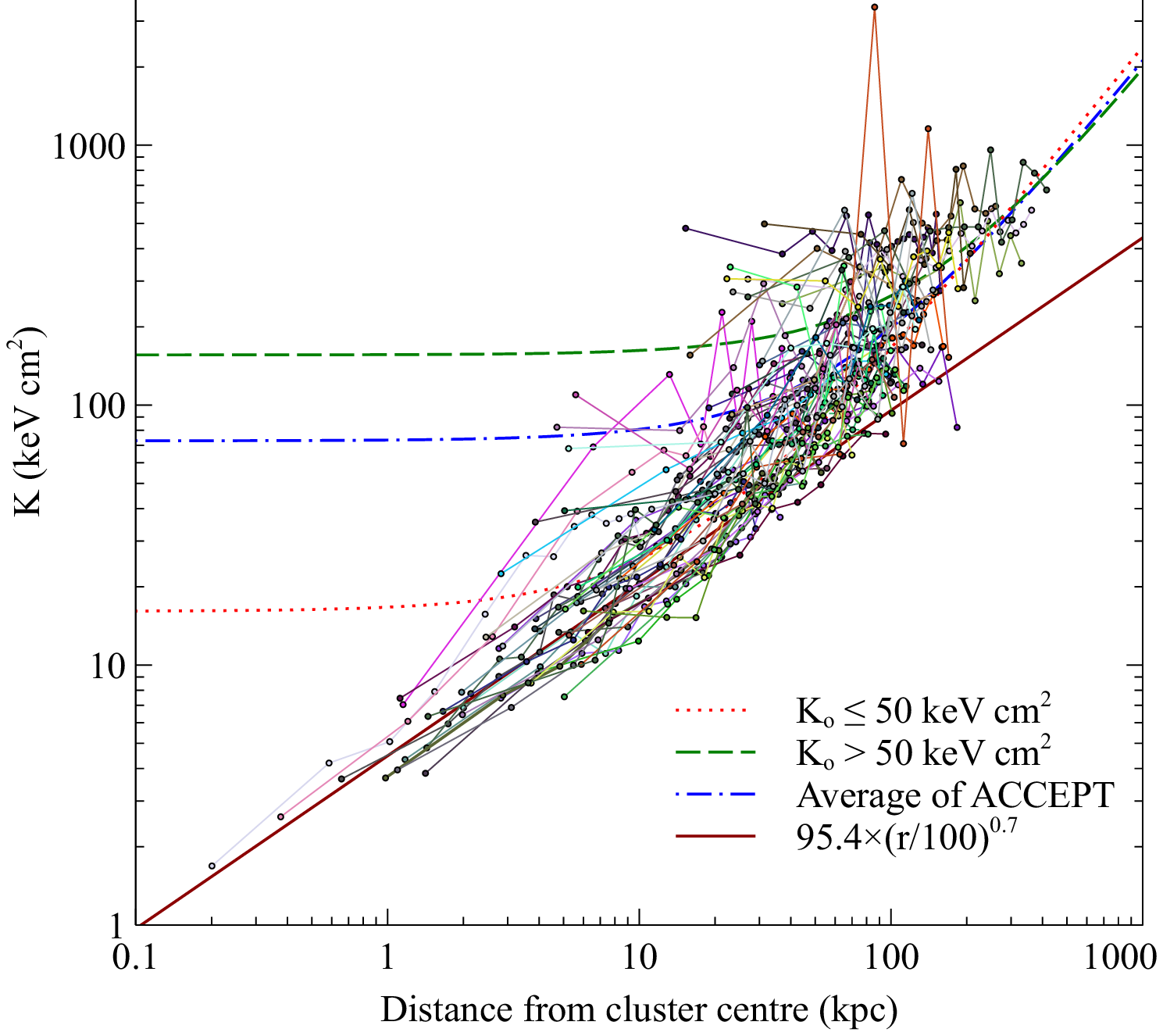}
\includegraphics[trim=0cm 14cm 4.5cm 0cm, clip, height=3.4in, width=3.4in]{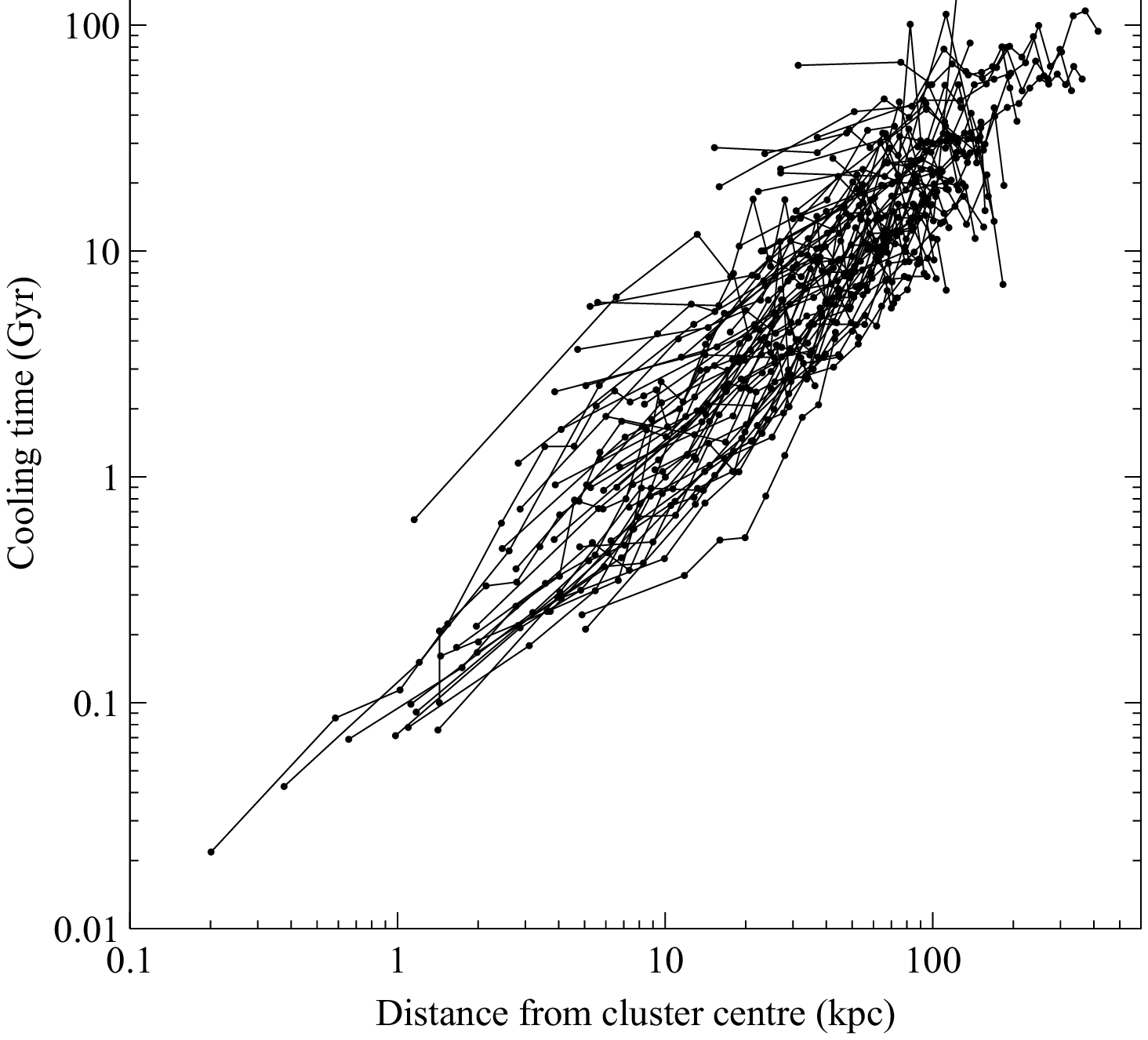}
\caption[]{{\it Left}: Overall entropy profile for 66 clusters from our sample. The green dashed line, the red dotted line and the blue dashed dotted line represent the mean best-fit entropy profiles for $K_{\rm o} >$ 50 keV cm$^{2}$, $K_{\rm o} \leq$ 50 keV cm$^{2}$, and all values of $K_{\rm o}$, from \cite{Cavagnolo09}, respectively. The brown line represents the best fit power law function, given by $K(r) = 95.4 \times (r/100 {\rm kpc}) ^{0.67}$. The sources used for this plot were randomly assigned different colours for their respective entropy profiles, to make individual profiles clearer. {\it Right}: Overall cooling time profile for the 66 clusters in our sample.}
\label{fig:entcool}
\end{figure*}

The main aim of our analysis is to study the radial temperature and number density profiles of the galaxy groups and clusters in our sample, in order to examine the shape of their overall entropy profile. To this end, we extracted spectra from a series of concentric annuli, which were centred on the peak of the X-ray emission of each cluster. For each cluster, the annular regions were defined in such a way so that they all contained the same number of counts, i.e. they all had the same signal-to-noise ratio. The total background-subtracted (if available), but not exposure-corrected, cluster images were used for this purpose, to ensure that the extracted spectra represented the same region in all datasets belonging to the same cluster. The number of annuli created for each cluster depended on the quality and quantity of data available for that particular cluster. The signal-to-noise ratio for the individual annuli of each source ranged from 28 to over 200.

%It is worth noting that some of the groups and clusters in our sample, such as Abell~2052 and Hydra~A, have clearly visible central point sources in their core. These central point sources were excised when the annular regions were created, in order to avoid contamination of the spectra from non-thermal emission. In all other cases, with few exceptions, the inner radius of the innermost annular region was $<$5kpc. To check whether these spectra were affected by non-thermal emission from the central galaxy, we fitted the spectra with  

In the case of the {\it Chandra} data, background spectra were extracted from the corresponding regions of the appropriate reprojected blank-sky background files. For the {\it XMM-Newton} spectra, a region free from group or cluster emission, and at a similar distance off-axis as the object of interest, was chosen to provide a local background during spectral analysis. The corresponding ancillary region files (ARFs), and redistribution matrix files (RMFs) were extracted using the {\sc ciao} {\sc mkwarf} and {\sc mkacisrmf} routines for {\it Chandra} data, and the {\sc sas} {\sc arfgen} and {\sc rmfgen} routines for the {\it XMM-Newton} data, respectively. To account for projection effects, the spectra were deprojected using the {\sc dsdeproj} routine, as described in \cite{Sanders07} and \cite{Russell08}. {\sc dsdeproj} is a model-independent deprojection method, which, assuming only spherical symmetry, removes projected emission in a string of shells. After deprojection, the spectra were binned to have a minimum of 25 counts in each spectral bin. As previously mentioned, we were unable to obtain deprojected radial temperature and electron number density profiles for all the groups and clusters in our sample. This is because of the low quality of the data available for these clusters, meaning that an accurate deprojection was not possible. 

It is worth noting that some of the groups and clusters in our sample, such as Abell~2052 and Hydra~A, have clearly visible central point sources in their core. These central point sources were excised when the annular regions were created, in order to avoid contamination of the spectra from non-thermal emission. In all other cases, with few exceptions, the inner radius of the innermost annular region was $<$5kpc. To check whether these spectra were affected by non-thermal emission from the central galaxy, we fitted the spectra with an additional power law component. In all cases, adding a power law component offered no improvement to the spectral fit. 

Spectral fits were performed using {\sc xspec} \citep{Arnaud96} v12.7.1b, in the 0.5--7.0 keV energy band, wherever possible. If it was not possible to perform spectral fits in this energy band, due to poor quality of the data towards higher ($>$5.0 keV) or lower ($<$0.7 keV) energies, we restricted ourselves to a more narrow energy range. If more than one dataset were available for a particular cluster, a joint spectral fit was performed between the corresponding spectra from each dataset. In the case of {\it XMM-Newton} data, joint spectral fits were also performed between the spectra from the different EPIC detectors used during the observation. The default model used for our spectral modelling was {\sc wabs*apec}, where the {\sc wabs} component (v 2.0.2) accounts for photoelectric absorption between us and the cluster, and the {\sc apec} component \citep{Smith01} models the thermal emission from the ICM. The available free parameters in our {\sc wabs*apec} model are the column density N$_{\rm H}$, the temperature of the ICM, k$_{\rm B}$T, the abundance of the ICM, the normalisation of the {\sc apec} component, which is effectively the emissivity measure of the ICM, and the redshift of the source. N$_{\rm H}$ was fixed at the Galactic value for each cluster \citep{Kalberla05}, while the redshift for each cluster was fixed to the value listed in the NASA Extragalactic Database (NED\footnote[1]{http://ned.ipac.caltech.edu}). The temperature, normalisation and abundance parameters of the {\sc apec} component were left free to vary. In certain cases, the quality of the data available for a cluster was too low for the abundance to be measured reliably. In these cases, the abundance was fixed to 0.3 Z$_{\odot}$. Where data of a high quality were available, it was possible to discern emission lines from elements other than iron, such as silicon, sulphur and magnesium. In these cases, we replaced the {\sc apec} component with a {\sc vapec} component, and allowed the appropriate abundances to vary, while fixing the helium abundance to solar and tying the rest of the abundances to that of iron. If more than one thermal component was needed to model the ICM emission in a certain region, an additional thermal component was added, and the abundances between the two thermal components were tied together.   

\subsection{Deprojected profiles}

Having obtained deprojected electron number density and temperature radial profiles for 66 of the clusters in our sample, we calculated entropy and cooling time profiles for each of these clusters. The definition used for the entropy was $K(r) = k_{\rm B} T \times n_{\rm e}^{-2/3}$. The formulae used to calculate the cooling time for each cluster were 
\begin{equation}
  t_{\rm {cool}} = 3.2\times10^{10} \rm{yr} (\frac{n_{e}}{10^{-3} \rm{cm}^{-3}})^{-1} (\frac{T}{10^{7} \rm{K}})^{0.5} , 
\label{eq:tcool1}
\end{equation}
for $T >$ 3$\times$10$^{7}$ keV, and
\begin{equation}
  t_{\rm {cool}} = 0.8\times10^{10} \rm{yr} (\frac{n_{e}}{10^{-3} \rm{cm}^{-3}})^{-1} (\frac{T}{10^{7} \rm{K}})^{1.6} , 
\label{eq:tcool2}
\end{equation}
for $T <$ 3$\times$10$^{7}$ keV \citep{Sarazin88}. The choice between the two formulae was done by studying the individual deprojected temperature profiles. As the electron number density and temperature measurements were made using the same spectral regions, there was no need to interpolate the deprojected number density or temperature profiles. 
%We note that Equations \ref{eq:tcool1} and \ref{eq:tcool2}, are not the most accurate equations used to calculate the cooling time of individual groups and clusters. This is because solar abundances are assumed, rather than the varying individual abundances being taken into account by means of the cooling function. However, they give us a good indication of the general trend of the cooling time profiles of the sources in our sample. 

The overall entropy and cooling time profiles are shown in the left hand and right hand panels of Figure \ref{fig:entcool}, respectively. The green dashed line, the red dotted line and the blue dashed dotted line in the entropy profile represent the mean best-fit entropy profiles for $K_{\rm o} >$ 50 keV cm$^{2}$, $K_{\rm o} \leq$ 50 keV cm$^{2}$ and all values of $K_{\rm o}$, from \cite{Cavagnolo09}, respectively. \cite{Cavagnolo09} define $K_{\rm o}$ as the excess of the core entropy above the best-fitting power law profile for the entropy at larger radii. They define the entropy $K(r)$ using the functional form
\begin{equation}
K(r) = K_{o}+K_{100}\times(r/100 {\rm kpc})^{\alpha}.
\label{eq:entcavagnolo}
\end{equation}
The sources used to create the overall entropy profile plot were randomly assigned different colours for their respective entropy profiles, to make individual profiles stand out more. As is obvious from our overall entropy profile, the flattening of the entropy profile towards smaller radii, that was suggested by \cite{Cavagnolo09}, is not observed in our sample. Instead, our entropy profiles can be modelled by a simple power law model, given by $K(r)=95.4\times(r/100 {\rm kpc})^{0.7}$, which is the best-fit power law model to our data, and is shown as a brown dotted line in the left hand panel of Figure \ref{fig:entcool}. Although the shape of the overall entropy profile is that of a power law, there are some individual sources whose entropy appears to flatten out towards the core. We point out that this ``flattening'' appears to take place mostly at radii $\geq$3--4 kpc, and is a result of the central spectral bin being relatively large, due to poor data quality.  

We note that Equations \ref{eq:tcool1} and \ref{eq:tcool2}, give only an approximate representation of the average cooling behaviour of the sources in our sample, and ignore any metallicity variations. The heating and cooling behaviour of the galaxy groups and clusters in our sample will be investigated in detail in future work. 

\begin{figure*}
\begin{center}
\includegraphics[trim = 0cm 14cm 4.5cm 0cm, clip, height=3.4in, width=3.4in]{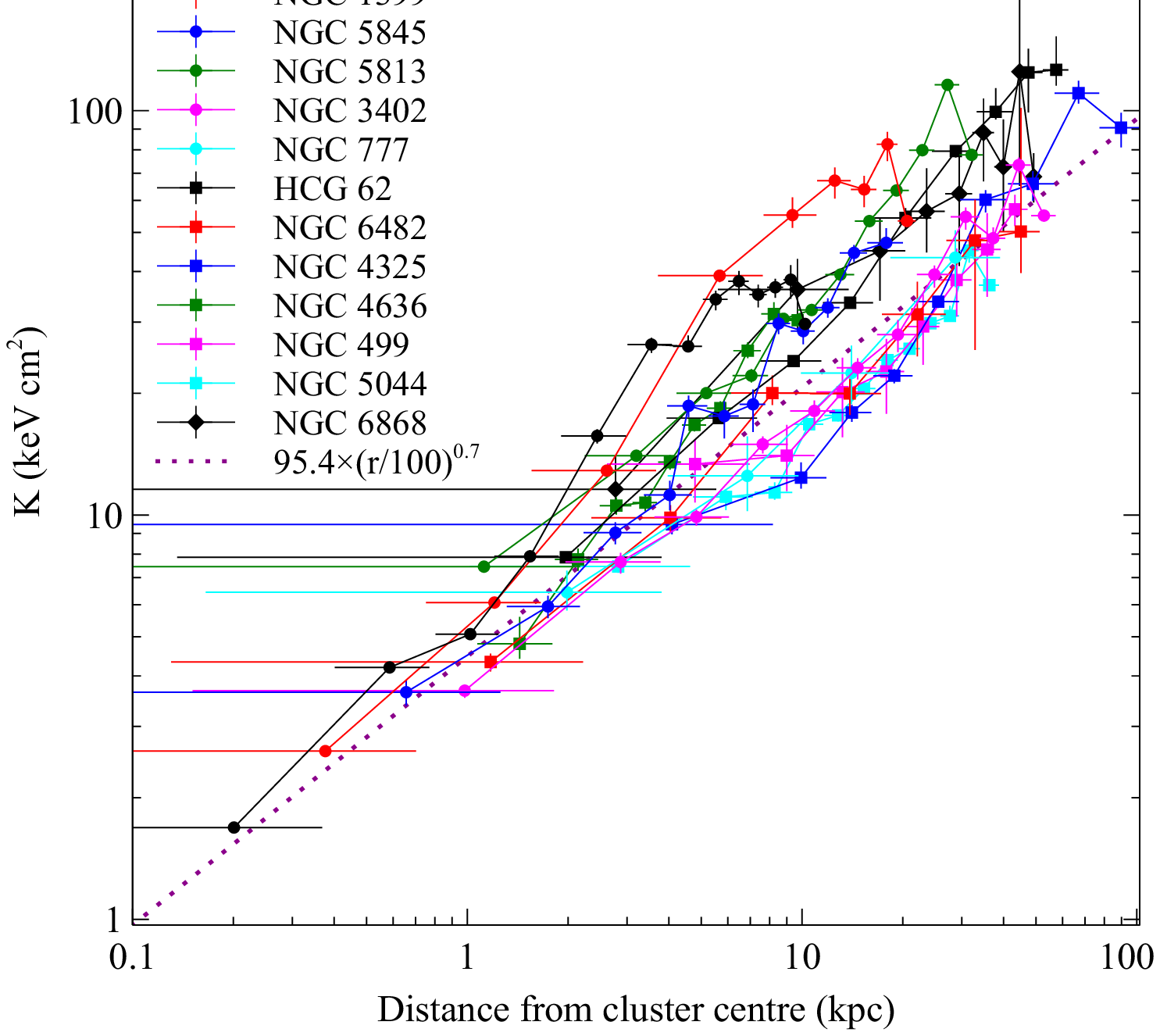}
\includegraphics[trim = 0cm 14cm 4.5cm 0cm, clip, height=3.4in, width=3.4in]{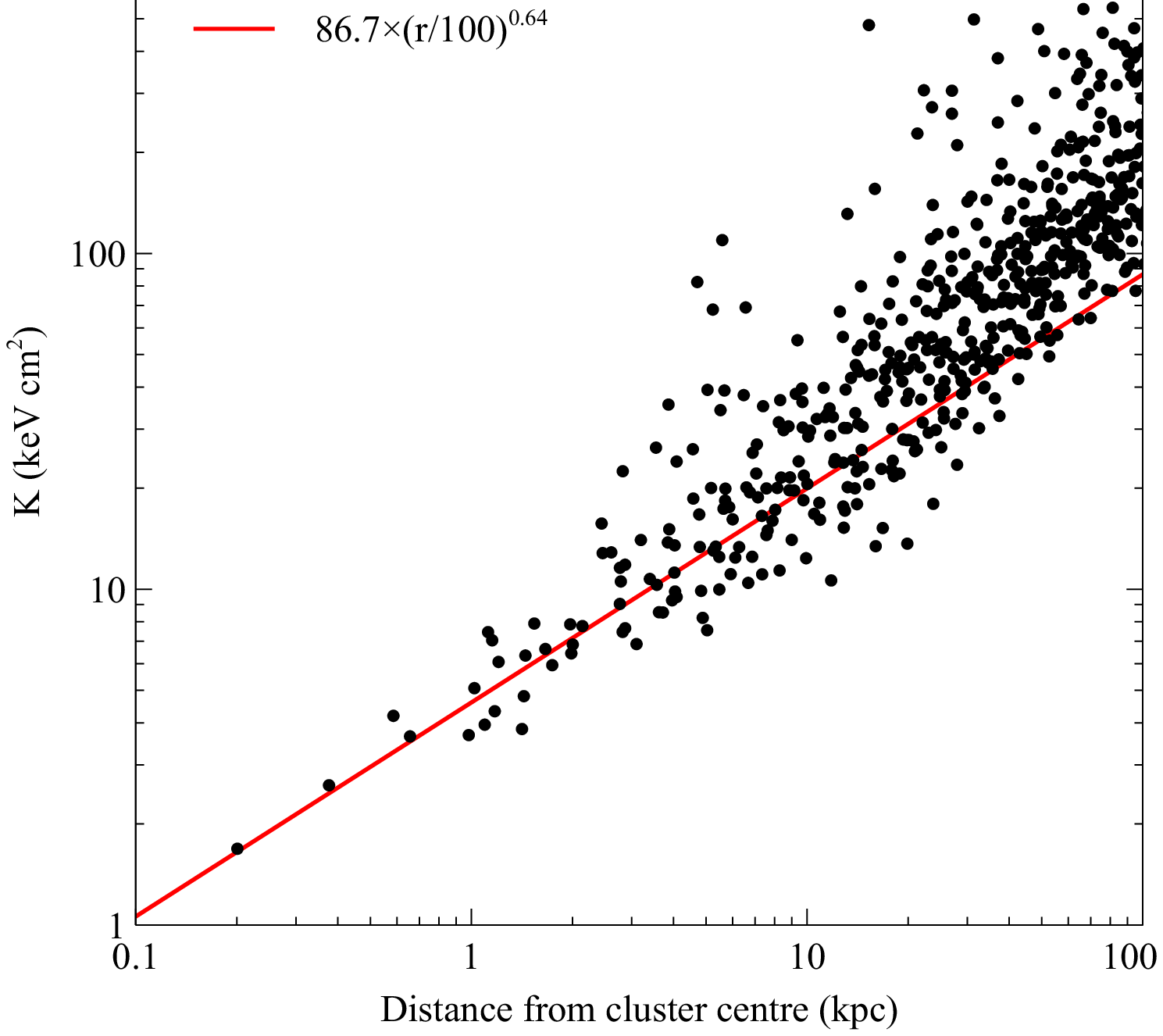}
\caption[]{{\it Left:} Entropy profiles for 13 sources with temperatures of $\leq$1.2 keV. The dotted line represents the best-fit power law model to the overall entropy profile. {\it Right:} The best-fit power law to the central 20 kpc of the overall entropy profile (red line) plotted over individual entropy profiles. Neither plot shows signs of the entropy flattening at small core radii, that are independent of the spatial resolution (i.e. the the size of the inner bin).}
\label{fig:13sources}
\end{center}
\end{figure*}

To further check whether the entropy profiles we calculated flatten at small radii or not, we focused on a subset of sources in our sample. We required these sources to have an innermost spectral bin centered at a distance of $\leq$5 kpc from the source centre, and, additionally, to have similar temperatures. This subset consisted of 13 groups and clusters, all of which have a temperature of $\leq$1.2 keV. Their entropy profiles are shown in the left hand panel of Figure \ref{fig:13sources}, where the dotted line represents the best-fit power law to the overall entropy profile. The data of these 13 sources are of a high quality: with the exception of NGC~777 and NGC~6482, whose annuli had a signal-to-noise ratio of 28 and 32 each, respectively, all the sources had annuli with a signal-to-noise ratio of 60 or more each. As can be seen, the entropy profiles show no sign of flattening towards the cores of these sources. 

In addition, we fitted the inner 20 kpc of the overall entropy profile with a power law model. The best fit power law model was given by $K(r)=86.7\times(r/100 {\rm kpc})^{0.64}$. We then overplot this model on the overall entropy profile, as shown in the right hand panel of Figure \ref{fig:13sources} (the aforementioned power law model is shown as a red line). Again, there appears to be no entropy floor at smaller radii. It is worth pointing out that most of the points in the right hand panel of Figure \ref{fig:13sources} lie above the red line. This indicates that the entropy profile could be represented by a broken power law, rather than a simple power law, out to larger radii. This is also suggested by the plot in the left hand panel of Figure \ref{fig:entcool}.

\subsection{A closer look at the Centaurus cluster}

In order to compare our results more closely with those of \cite{Cavagnolo09}, we selected one cluster, present in both samples, for which we obtained radial temperature, electron number density and entropy profiles. We choose to use NGC~4696, the brightest cluster galaxy of the Centaurus cluster. First of all, we extracted spectra for this cluster in 22 concentric annular bins, each of which contained 115000$\pm$5000 counts. To account for projection effects, the spectra were deprojected using the {\sc dsdeproj} routine. These spectra were jointly modelled using a {\sc wabs*vapec} model in {\sc xspec} v12.7.1b. The free parameters in our model were the temperature, the normalization and the iron, silicon and sulphur abundances of the {\sc vapec} component. The column density was fixed at 8.31$\times$10$^{20}$ cm$^{-2}$, while the helium abundance was fixed at solar. The remaining abundances were tied to that of iron, and the individual parameters of the spectra from different datasets that corresponded to the same spectral region were tied together. 

As the temperature structure in the core of NGC~4696 is bimodal \citep{Sanders02, Sanders08, Panagoulia13}, we performed an additional fit of the two central spectral bins ($\sim$5 kpc), with a two-temperature model, namely {\sc wabs*(vapec+vapec)}. In this case, the spectra from the two bins were jointly fit, with all the free parameters, except the temperature and normalisation of each {\sc vapec} component, tied together. The free parameters in this model were the temperatures, normalizations, the column density, and the oxygen, neon, magnesium, silicon, sulphur, calcium, iron and nickel abundances. The argon abundance was tied to that of calcium, while the carbon, nitrogen and aluminium abundances were fixed at 0.3 Z$_{\odot}$. The entropy for each of the gas components in the two annuli was calculated assuming pressure equilibrium between the gas components of each annulus. 

In addition to the spectral deprojection method, we also used a surface brightness (SB) profile to obtain a radial entropy profile of a higher spatial resolution. For this reason, we used 3 times as many SB bins as spectral bins. Each SB bin contained 41500$\pm$5000 counts. After extracting an SB profile, this was then deprojected \citep[following][]{Fabian81} assuming an NFW potential \citep{Navarro96}. For more details on the algorithm used, see Section 3.1 in \cite{Sanders10}. From the resulting set of potential, pressure, temperature and density profiles, obtained from the SB deprojection, an entropy profile can be calculated. 

\begin{figure}
\includegraphics[trim = 0cm 14cm 4.5cm 0cm, clip, height=3.4in, width=3.4in]{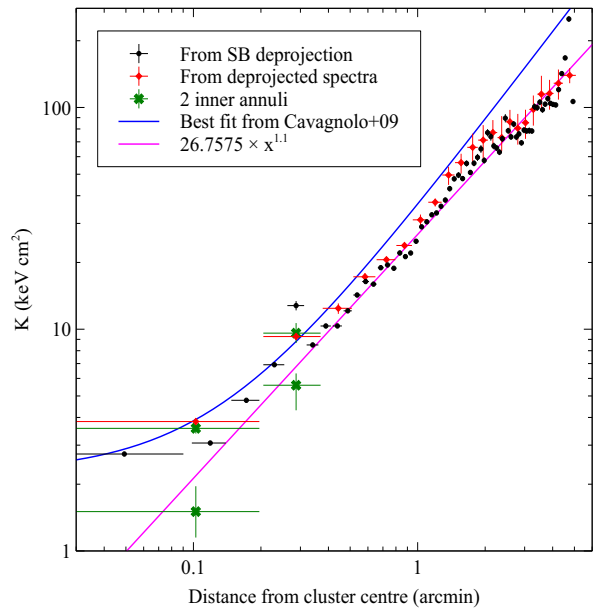}
\caption[]{Entropy profiles for NGC~4696. The black circles, red diamonds and green crosses represent the entropy profiles obtained from the SB deprojection, the spectral deprojection and the two-temperature fit of the two inner spectral bins, respectively. The blue line indicates the best-fit entropy profile from \cite{Cavagnolo09}, while the pink line indicates a power law fit to the SB deprojection entropy profile.}
\label{fig:cenent}
\end{figure}

The resulting entropy profiles are shown in Figure \ref{fig:cenent}. The black circles, red diamonds and green crosses represent the entropy profiles obtained from the SB deprojection, the spectral deprojection and the two-temperature fit of the two inner spectral bins, respectively. The blue line indicates the best-fit entropy profile from \cite{Cavagnolo09}, while the pink line indicates a power law fit to the SB deprojection entropy profile. The two methods used in this work to obtain the entropy profile are in good agreement with each other, while being offset from the best fitting profile of \cite{Cavagnolo09}. This is due to the fact that \cite{Cavagnolo09} used projected temperature profiles to calculate entropy profiles, while our analysis uses deprojected temperature profiles. We note that the entropy value of one of the points in the SB deprojection entropy profile, namely the one at 0.3 arcmin, is quite a bit higher than the value calculated using spectral deprojection, and the mean profile from \cite{Cavagnolo09}. This is due to the fact that this point at 0.3 arcmin lies in the middle of the radio bubbles in NGC~4696, where pressure is also provided by the non-thermal material in the bubbles. As a result, the SB deprojection compensates for the apparent lack of pressure by overestimating the gas temperature. In addition, the assumption of spherical symmetry implemented by the SB deprojection algorithm, is not as good an approximation in the annuli in which the bubbles reside.  

Figure \ref{fig:cenent} also shows how entropy measurements can be affected, if a complex temperature structure is present, and more than one thermal component is needed to model the emission from that region. As can be seen, there are two different components to the entropy in the central 2 spectral bins (green crosses), straddling either side of the power law fit to the SB deprojected entropy profile (pink line). If the two entropy components are taken into account, there is no visible flattening of the entropy profile at small radii.  We note that the resulting entropy profile from the SB deprojection that we performed is similar to the entropy profile in \cite{Cavagnolo09} at small radii, since in both cases, the gas is assumed to be single-phase. This example highlights the difference in results that is caused by using different methods in obtaining entropy profiles. 

%Need to mention that the gas follows a power law at larger radii, hence it would make sense that it should follow a power law at smaller radii. 

\section{Discussion}
\subsection{Core entropy flattening: A resolution effect?}

The flattening of the entropy profile towards the cluster core, seen by \cite{Cavagnolo09} (see Figure 5 in the same paper), could be a resolution effect. \cite{Cavagnolo09} use a minimum of 3 spectral bins, with a minimum of 2500 counts each, to obtain projected radial temperature profiles for each cluster. To obtain electron number density profiles, they used an exposure-corrected, background-subtracted and point-source cleaned surface brightness profile, from which they obtained emission measure profiles in concentric annuli with a width of 5'' each. The surface brightness profiles were then deprojected, using the corresponding spectral count rates and normalizations, which were obtained by interpolating the radial temperature profile on to the surface brightness radial grid. This allowed the surface brightness profile to be converted into a deprojected electron number density profile. As a result, when the entropy profiles were calculated, multiple number density bins were covered by one single temperature bin. This is a potentially significant problem for cluster cores, where the temperature structure is expected to be complex. To tackle this problem, \cite{Cavagnolo09} assumed that the temperature remained constant in the central temperature bin, as well as the electron number density bins that were covered by it. As a result, an existing, real temperature gradient in the core of a cluster could be overlooked, leading to a less steep entropy profile in the core.

\begin{figure}
\includegraphics[trim = 0cm 14cm 5cm 0cm, clip, width=3.4in, height=3.4in]{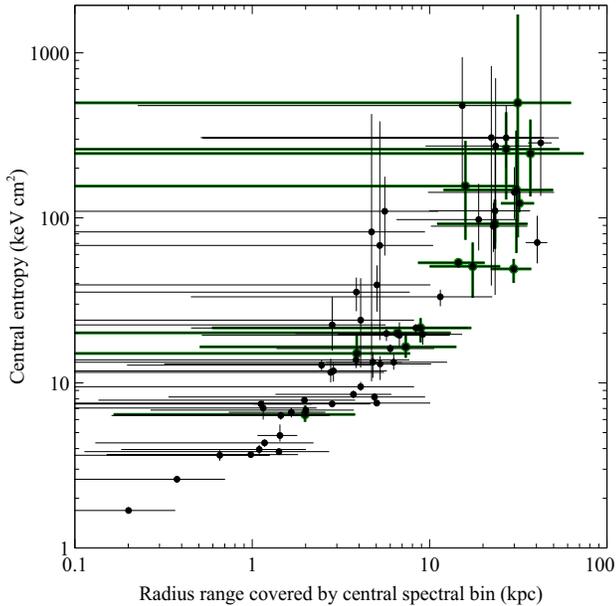}
\caption[]{Calculated central entropy vs central spectral bin size for the 66 groups and clusters used to calculate the overall entropy profile. The points outlined in green are the sources for which only {\it XMM-Newton} data, or less than 15 ks of {\it Chandra} data, were available. There is clearly a correlation between the central bin size and the calculated value for the central entropy.}
\label{fig:centralentvsbinsize}
\end{figure} 

In addition, the exposure times of some of the observations used by \cite{Cavagnolo09} in their analysis are rather short. The fact that they placed a lower limit of 2500 counts per spectral bin when extracting temperature profiles, could have lead to the creation of rather large central spectral bins. Once again, this means that an existing temperature gradient in a cluster core may have been smoothed out, leading to an ``artificial'' flattening of the entropy profile. The use of projected, rather than deprojected, temperature profiles by the same authors augments the amount by which a temperature gradient is smoothed out. This is because projected profiles include emission from neighbouring regions, not only the region of interest. 

%\subsection{Implications for heating}

The existence of an entropy ``floor'' at small radii depends on the assumption that large amounts of gas are deposited there by heating mechanisms, both at the present time and in the past. However, cooling flows transport large amounts of cool, dense gas to the central regions of galaxy clusters. 
%We therefore expect to see a power-law entropy profile right down to the cluster centre, for at least all the clusters in which heating is not sufficient to offset the cooling flow. 
It is clear that a heating-cooling balance has been struck in these objects. The appearance of entropy floors across a wide range of redshifts in the work of \cite{McDonald13}, is hard to explain. We note that the clusters in the study of \cite{McDonald13} are distant and have low levels of counts, while our sources are closer, and so far better resolved.

To demonstrate that the detection of an entropy floor can be a resolution effect, we plot the calculated value for the entropy in the central spectral bin of each cluster, vs the radius range covered by the central spectral bin. The resulting distribution is shown in Figure \ref{fig:centralentvsbinsize}. The points outlined in green are the sources for which only {\it XMM-Newton} data, or less than 15 ks of {\it Chandra} data, were available. It is obvious that the measured central entropy is correlated with the central spectral bin size: the larger the central spectral bin, the larger the value measured for the central entropy. This may be because larger spectral bins contain emission from larger volumes of a group or cluster, meaning that any temperature and number density gradients are smoothed out. This figure also demonstrates that poor data quality, or poor spatial resolution, can indeed lead to the miscalculation of the central entropy of a galaxy group or cluster.  

\begin{figure}
\includegraphics[trim = 0cm 14cm 5cm 0cm, clip, width=3.4in, height=3.4in]{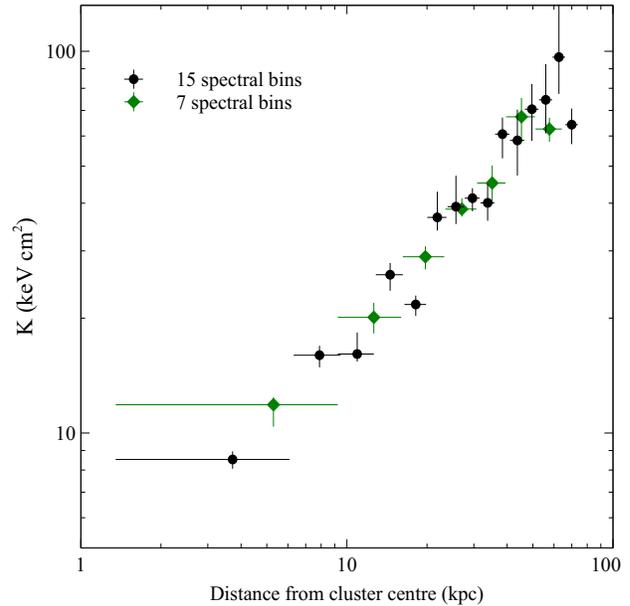}
\caption[]{Entropy profiles for A~3581. The black circles represent the entropy profile calculated using 15 spectral bins, and is the same one as used in the left hand panel of Figure \ref{fig:entcool}. The green diamonds indicate the new entropy profile, which was recalculated using 7 larger spectral bins. In the new entropy profile, the central entropy is increased by $\sim$40 per cent with respect to the original entropy profile.}
\label{fig:a3581}
\end{figure}

To determine the effect of using larger spectral bins, and therefore smoothing any existing temperature and number density gradients, we derive the entropy profile for one of our clusters, A~3581, using spectral bins twice as wide as those used in our original entropy calculations. This reduces the number of spectral bins from 15 to 7, though the signal-to-noise ratio increases from 122 to 173. The calculation of the new entropy profile was done as described in Sections 4.1 and 4.2. Both the old and new entropy profiles are shown in Figure \ref{fig:a3581}. Although the larger spectral bins do not affect the general shape of the entropy profile, the central entropy has increased by about 40 percent, from 8.5 keV cm$^{2}$ to 11.9 keV cm$^{2}$. We point out that A~3581 is a cluster for which high-quality {\it Chandra} data are available, so just using larger spectral bins is not likely to change the overall shape of the entropy profile. However, this might be a significant effect for sources with data of a poorer quality, where the inner regions are undersampled. 

We note that there is indeed a population of clusters that do have large central entropies, which are non-CC clusters. One such cluster is the Coma cluster, which we calculate to have a central entropy value of $\sim$480 keV cm$^{2}$. There is, however, low entropy gas in the Coma cluster, surrounding the two BCGs, as found by \cite{Vikhlinin01}. In fact, X-ray coronae of early-type galaxies embedded in hot cluster atmospheres seem to be relatively common, having withstood processes such as ICM stripping and intense cooling, and their properties have been studied in detail by e.g. \cite{Sun07}.  

\subsection{Central entropy distribution}

\begin{figure}
\includegraphics[trim = 0cm 14cm 5cm 0cm, clip, width=3.4in, height=3.4in]{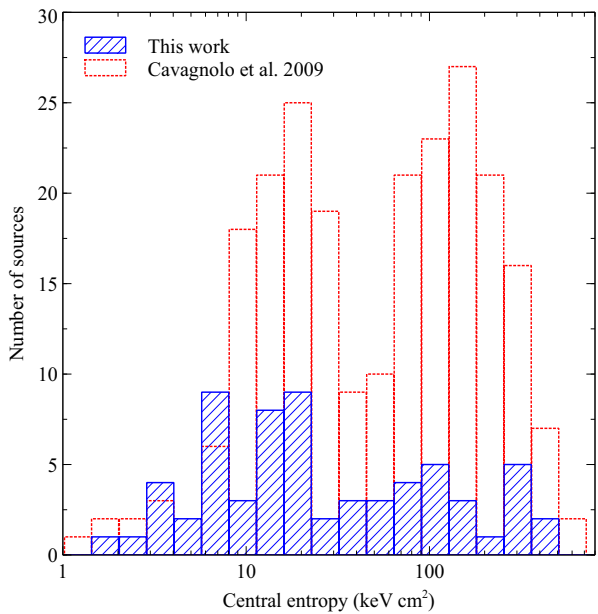}
\caption[]{Distribution of central entropies calculated in this work (blue shaded histogram), overlaid with the distribution of core entropies reported in \cite{Cavagnolo09} (red dotted histogram). The width of the bins is 0.15 in log space. We see a broad single-peaked distribution rather than a bimodal distribution. The peak in our distribution coincides with the first peak of the \cite{Cavagnolo09} distribution.}
\label{fig:coreent}
\end{figure}

The entropy of a gas is directly related to the amount of time it takes it to lose its energy through radiative cooling. Low entropy gas cools radiatively at higher rates than high entropy gas. The distribution of central entropies can therefore help us undestand more about the physical processes in cluster cores, such as AGN heating, as well as the timescales on which they operate. For this reason, we plot the logarithmically binned distribution of the central entropies in Figure \ref{fig:coreent} (blue shaded histogram). The bin widths are 0.15 in log space. We have also overplotted the distribution of core entropies shown in \cite{Cavagnolo09}, as the red dotted histogram. We observe a peak in the distribution at $\sim$15 keV cm$^{2}$, and the distribution then flattens out smoothly. We do not observe the double-peaked distribution reported in \cite{Cavagnolo09}, though we note that the first peak of the \cite{Cavagnolo09} distribution coincides with the peak of our distribution. 

We point out that, while the values for $K_{o}$ obtained in \cite{Cavagnolo09} are the result of fitting a functional form for the entropy (Equation \ref{eq:entcavagnolo}) to individual entropy profiles, the values for the central entropy that we obtain are the result of performing spectral fits to the innermost spectral bins. Therefore, our central entropy values can be viewed as upper limits, and are limited by the spatial resolution and quality of the data. 

As is visible in Figure \ref{fig:centralentvsbinsize}, 50 per cent of our sources with central entropies $\geq$ 30 keV cm$^{2}$ have data of a relatively poor quality and/or low spatial resolution. This means that the shape of the high-entropy tail in our distribution might look different in reality from the one plotted. The reason for this is that we cannot be certain of the position of some of our sources in our distribution, due to poor data quality. This may also explain the second peak in the distribution of \cite{Cavagnolo09}. 

We note that our sample is smaller than that of \cite{Cavagnolo09}, and consists of nearby clusters. Hence, if the bimodal distribution of \cite{Cavagnolo09} is indeed real and due to a combination of episodic AGN feedback and merger activity, we would expect to see a similar distribution in our analysis. The fact that we do not observe this, suggests that AGN feedback may be better regulated and less episodic than previously believed. 

\begin{figure}
\includegraphics[trim = 0cm 14cm 5cm 0cm, clip, width=3.4in, height=3.4in]{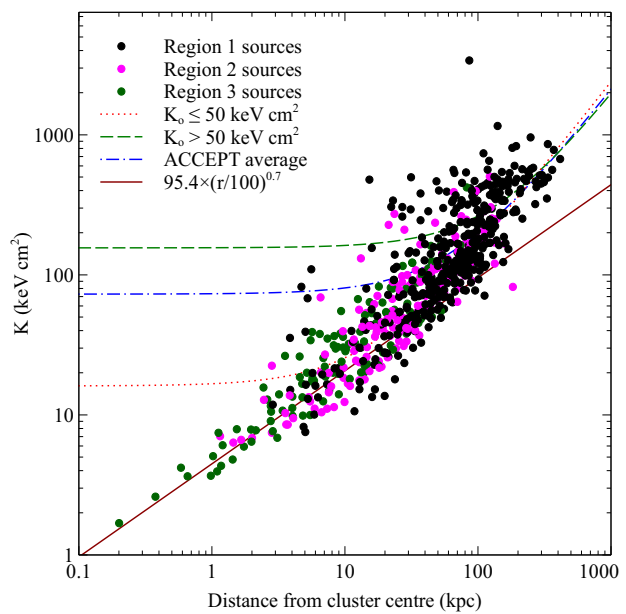}
\caption[]{Overall entropy profile for the 66 sources in our sample, colour-coded for sources from different $z-L_{X}$ regions (see Section 2). Region 1 sources are represented by black circles, Region 2 sources by pink circles, and Region 3 sources by green circles. The lines in the plot are the same as those in the left-hand panel of Figure \ref{fig:entcool}.}
\label{fig:entcolour}
\end{figure}

In order to check the relative contributions of sources from the 3 different $z$-$L_{X}$ regions (see Section 2), we replot the overall entropy profile, this time colour-coding the sources from the same region. The resulting overall entropy profile is shown in Figure \ref{fig:entcolour}. The black circles indicate Region 1 sources, the pink ones Region 2 sources, and the green ones Region 3 sources. The lines in the plot are the same as in the left panel of Figure \ref{fig:entcool}. As expected, the highest entropy values are dominated by Region 1 and 2 sources, which are the most luminous, hotter clusters in our sample, and are generally further away, so having relatively poorly resolved cores. On the other hand, Region 2, and particularly Region 3, sources populate the lowest entropy values. This is what we would expect, as Region 2 and 3 sources consist of nearby less-luminous, and therefore cooler, groups and clusters. 

After inspecting the images in our sample, we determine that 9--10 per cent of the sources in our sample are undergoing a merger, or have done so recently. Half of these sources have a central entropy value $>$110 keV cm$^{2}$, which could indicate that the merging activity has indeed quenched the cooling flow in the larger of the merging substructures. 

\subsection{Bondi accretion}
The lack of an entropy floor in the core of galaxy clusters could have an impact on the plausibility, and importance, of Bondi accretion fuelling the black hole in the central galaxy \citep[see e.g][where the first study argues for Bondi accretion, and the other two against]{Allen06, Rafferty06, McNamara11}. The Bondi accretion rate is given by
\begin{equation}
\dot{M}_{\rm Bondi} = 4 \pi R_{\rm Bondi}^{2} \rho c_{\rm S} ,
\end {equation}
where $\rho$ is the ambient density, $c_{\rm S}$ is the sound speed in the medium, and $R_{\rm Bondi}$ is the Bondi radius, defined as
\begin{equation}
R_{\rm Bondi} = \frac{G M}{c_{\rm S}}^{2} , 
\end{equation}
where $M$ is the mass of the black hole. Therefore, $\dot{M}_{\rm Bondi}$ can be written as 
\begin{equation}
\dot{M}_{\rm Bondi} = 4 \pi \frac{(G M)^{2}}{c_{\rm S}^{3}} \rho \propto M^{2} \rho T^{-3/2}.
\end{equation}
%The density $\rho$ is given by $\rho = n \times m_{\rm p}$, where $n$ is the number density of the medium and $m_{\rm p}$ is the mass of a proton. In addition, the sound speed $c_{\rm S}$ can be expressed as
%\begin{equation}
%c_{\rm S}^{2} = \frac{k_{\rm B} T}{m_{\rm p}} , 
%\end{equation}
%so we have 
%\begin{equation}
%\dot{M}_{\rm Bondi} = 4 \pi \frac{(G M)^{2}}{(k_{\rm B} T)^{3/2}} n m_{\rm p} . 
%\end{equation}
Making the substitution $P = n k_{\rm B} T$, where $P$ is the pressure of the medium, the Bondi accretion rate can be written as 
\begin{equation}
\dot{M}_{\rm Bondi} \propto M^{2} P T^{-5/2}. 
\end{equation}
Alternatively, we can use entropy so that 
%\begin{equation}
%K^{3/2} = \frac{T^{3/2}}{n_{\rm {e}}}.
%\end{equation}
%Equation (7) can then be rewritten as
\begin{equation}
\dot{M}_{\rm Bondi} \propto M^{2} K^{-3/2}.
\end{equation}

There is clearly a strong dependence of $\dot{M}_{\rm Bondi}$ on temperature and/or entropy. Therefore, if the core entropy of a core cluster does not flatten out, and is, for example, half the value assumed from a flat core entropy profile, this greatly impacts the Bondi accretion rate of the AGN. A drop of a factor of 2 temperature could increase the Bondi accretion rate by a factor of $\sim$6. Alternatively, a factor of 10 drop in entropy leads to a factor of $\sim$31 increase in $\dot{M}_{\rm Bondi}$.

\subsection{Feeding the AGN}
The inner entropy profile is thus clearly important for the viability of hot mode accretion fuelling the central black hole. Our results improve the chance that it is important. Whether it is sufficient is difficult to assess, because the Bondi radius is unresolved in all but the closest objects \citep[M87, NGC~3115, and Sgr~A* in][respectively]{diMatteo03, Wong13, Wang13}. Inward extrapolation of measured temperature and density profiles is possible, but has led to conflicting results. For example, \cite{Allen06} find that AGN jet power correlates with Bondi accretion rate, but \cite{Russell13} find a weaker correlation. The alternative is cold-mode accretion, where the inner cold gas clouds, commonly seen in cool core cluster BCGs \citep[e.g.][]{Werner13b}, are assumed to accrete \citep[e.g.][]{Soker08, McNamara12}. Cloud-cloud collisions are often invoked to overcome the expected angular momentum barrier to such an inflow. Current data, again, have insufficient spatial resolution to test this hypothesis. 

Other work assumed that the inner hot phase cools to produce cold clouds, which together make up the accretion flow \citep[e.g.][]{Gaspari13, Guo13, Li12}. A problem here is that the observed cool clouds tend to be dusty and molecular \citep[again, see][]{Werner13b}, which is difficult to explain if the gas is dust-free before cooling. Moreover, the expected soft X-ray emission lines expected from cooling gas are either not observed, or are very weak \citep[e.g. O VII,][]{Sanders11}. As mentioned in that paper and in \cite{Werner13a}, photoelectric absorption by cooled gas may, however, obscure such emission. The complexity and multiphasedness of the centre of the well-observed Centaurus cluster centred on NGC~4696 \citep{Panagoulia13}, emphasises that any clear solution to the hot-mode vs cold-mode debate will not be simple.  

Gas does pass from the hot phase to the cold phase, probably through a combination of radiative cooling and mixing \citep[see the discussion in][]{Fabian12}. The high observed incidence of bubbles in clusters and groups with short radiative cooling times means that the duty time of the bubbling action is high, so the inner regions are likely continuously stirred by this action. Much of the action may be in the form of outward dragging of gas, which further complicates the determination of the motion close to the central black hole. 

\subsection{X-ray cavity dynamics}
Since our sample covers a wide range of cluster morphologies, some of the sources in our sample display one or more sets of X-ray cavities. Of a particular interest is the subset of groups and clusters which have central cooling times $\leq$1 Gyr, which are much shorter than the age of the respective groups and clusters. The gas in the core of these groups and clusters is closely linked to the cooling rate in the respective group or cluster, and provides a unique insight into AGN feedback. This subset can also be used to study the relative prevalence of heating and cooling mechanisms in different types of groups and clusters, and whether there are any fundamental differences in the way they operate in groups and clusters. \\
There are 25 groups and clusters in our sample with a central cooling time $\leq$1 Gyr. The detailed study of this subset of objects, and in particular of their X-ray cavities, will be presented in a future paper. \\

\section{Summary}
Using the NORAS and REFLEX cluster catalogues, we created a volume- and $L_{\rm X}$-limited sample of 101 galaxy groups and clusters, all of which had archival {\it Chandra} and/or {\it XMM-Newton} data. We obtained deprojected radial number density and temperature profiles for all the sources in our sample. Out of the 101 groups and clusters in our sample, 66 sources had data of sufficient quality to provide reliable deprojected temperature and number density profiles. These sources were then used to create an overall cooling time and entropy profile. The main results of our analysis are:
\begin{itemize}
  \item{We do not observe flattening of the entropy profile towards the core of the sources in our sample, as has been reported in previous studies, e.g. \cite{Cavagnolo09}. In contrast, a power law profile of the form $K(r) = 241.2 \times (r/100 {\rm kpc}) ^{0.9}$ appears to model our overall entropy profile well. In fact, our analysis suggests that our overall entropy profile may be best described by a broken power law model.} 
      \item{By analysing the entropy profile of the Centaurus cluster in greater detail, we find that the existence of multi-temperature gas in the core of a cluster can affect entropy measurements. By fitting a two-temperature model to the central 5 kpc of the Centaurus cluster, we show that the entropy does, in fact, continue to decrease, rather than flatten out.} 
          \item{The flattening of the entropy profile reported in previous studies may be a resolution effect. The use of data of relatively poor quality, in combination with the analysis methods used, means that a temperature drop in the core of a cluster may have been smoothed out. In addition, if different methods are used to obtain the radial temperature and number density profiles, this can again lead to a temperature gradient in the cluster core being overlooked. Both these effects can result in the calculation of a ``false'' entropy floor in the core of a cluster.}
          \item{The non-existence of an entropy floor in the core of galaxy clusters can have an effect on the plausibility, and significance, of Bondi accretion in AGN. If, for example, the actual entropy in the core of a cluster was a tenth of the value of the entropy floor, this would imply a factor of $\sim$30 increase in the Bondi accretion rate.}
\end{itemize}

Work is currently underway to study the dynamics of the X-ray cavities of a subset of our group and cluster sample, namely 25 sources that have a central cooling time $\leq$1 Gyr. This will allow us to enhance our understanding of the AGN feedback cycle in low-redshift galaxy groups and clusters. 

\section*{Acknowledgements}
EKP acknowledges the support of a STFC sudentship. ACF thanks the Royal Society for support.

The plots in this paper were created using {\sc veusz}.\footnote{http://home.gna.org/veusz/}
The NASA Extragalactic Database (NED)\footnote{http://ned.ipac.caltech.edu/} was used to obtain redshifts for the spectral analysis presented in this work. 

\bibliographystyle{mn2e}
\bibliography{references}

\end{document}